\documentclass[twocolumn,floatfix]{revtex4}

\usepackage{amsmath}
\usepackage{graphicx}
\usepackage{amssymb}

\begin{document}

\preprint{As of 21th of January.}

\title{Nonlinear dynamics of Bose-condensed gases by means of a
low- to high-density variational approach}

\author{Alexandru I. Nicolin}
\email{nicolin@nbi.dk}
\affiliation{Niels Bohr Institute, Blegdamsvej 17, Copenhagen, 2100-DK}

\author{R. Carretero-Gonz\'alez}
\homepage{http://rohan.sdsu.edu/~rcarrete/}
\affiliation{Nonlinear Dynamical Systems Group%
\footnote{URL: {\tt http://nlds.sdsu.edu}},
Department of Mathematics and Statistics,
and Computational Science Research Center,
San Diego State University, San Diego CA, 92182-7720, USA}

\begin{abstract}
We propose a versatile variational method to investigate the
spatio-temporal dynamics of one-dimensional magnetically-trapped
Bose-condensed gases. To this end we employ a \emph{q}-Gaussian
trial wave-function that interpolates between the low- and the
high-density limit of the ground state of a Bose-condensed gas.
Our main result consists of reducing the Gross-Pitaevskii
equation, a nonlinear partial differential equation describing the
$T=0$ dynamics of the condensate, to a set of only three
equations: \emph{two coupled nonlinear ordinary differential
equations} describing the phase and the curvature of the
wave-function and \emph{a separate algebraic equation} yielding
the generalized width. Our equations recover those of the usual
Gaussian variational approach (in the low-density regime), and the
hydrodynamic equations that describe the high-density regime.
Finally, we show a detailed comparison between the numerical
results of our equations and those of the original
Gross-Pitaevskii equation.
\end{abstract}

\pacs{05.45.Xt, 03.75.Nt}

\maketitle

\section{introduction}

Due to the unprecedented experimental maneuverability of
Bose-condensed gases, their intrinsically nonlinear dynamics
became an appealing research topic engrossing scientists from
fields as diverse as numerical analysis and condensed matter
physics \cite{BEC-book,Boris-book,RCGbook}. The $T=0$ dynamics of
these gases is accurately described by the so-called
Gross-Pitaevskii equation (GPE), a cubic Schr\"odinger equation
brought forward in the early sixties by two seminal papers of
Gross and Pitaevskii \cite{pitaevskii,gross}.

While there is good quantitative agreement between the full GPE
and current experimental results on a wide range of topics,
including: dark soliton dynamics inside maganetic traps
\cite{expd,Denschlag2000a}, bright soliton interactions
\cite{strecker:prl:2002}, non-equilibrium oscillations in binary
BECs \cite{rcg:58}, and Faraday waves in periodically driven BECs
\cite{faraday}, there are, however, rather few models giving
analytical insight. Naturally, reducing the partial-differential,
therefore \emph{infinitely-dimensional}, GPE to the level of a few
ordinary differential equations (ODEs) is a nontrivial task.
Previous works include the so-called Gaussian variational model
(which describes the low-density regime)
\cite{Voptics,Boris_Prog,Zoller-var,ripoll-perez-garcia} and the
hydrodynamic analysis of high-density condensates
\cite{stringari,dalfovo_one}. All these works addressed the
dynamics of a Bose-condensed gas close to its ground state.
For a comprehensive account, from both an experimental
and theoretical viewpoint, of excited states such as
solitons, vortices and alike we refer to the recent
review \cite{BECBOOK} and references therein.
In this paper we present an efficient universal variational model
able to describe both the high- and the low-density dynamics close
to the ground state (\emph{i.e.}, without entering the realm of
soliton-like solutions). Our model stems from the
\emph{q}-Gaussian trial wave-functions that were used to describe
the ground-state properties of a trapped Bose-condensed gases
\cite{Fetter,Lenzi,Erdemir}. Though restricted to ground-state
properties, the work of Fetter \cite{Fetter} is particularly
relevant since he is the first one to point out the universality
of this ansatz.

The rest of the paper is structured as follows: In
Sec.~\ref{sec:VA} we present the variational method using the
\emph{q}-Gaussian trial wave-function. In Sec.~\ref{sec:VA_coup}
we reduce the variational equations to the level of three coupled
ODEs and in Sec.~\ref{sec:VA_decoup} we simplify these equations
to the level of two coupled nonlinear ordinary differential
equations describing the phase and the curvature of the
wave-function, and a separate algebraic equation yielding the
generalized width. In Secs.~\ref{sec:high} and \ref{sec:low} we
show that our equations recover those of the variational Gaussian
model and the standard hydrodynamic formulation of the GPE.
Section \ref{sec:num} presents our numerical results for a
particular case of a GPE with a periodically forced nonlinearity.
Finally, Sec.~\ref{sec:conclu} is alloted to our concluding
remarks.

\section{variational model\label{sec:VA}}

\subsection{Coupled equations\label{sec:VA_coup}}

The equation describing the $T=0$ dynamics of a trapped
three-dimensional Bose-condensed gas is given, in adimensional
units, by
\begin{equation}
i\frac{\partial\psi(\textrm{{\bf r}},t)}{\partial t}=\left(-\frac{1}{2}\nabla^{2}+V(\textrm{{\bf r}})+U(t)\left|\psi(\textrm{{\bf r}},t)\right|^{2}\right)\psi(\textrm{{\bf r}},t),\label{GPE_first}
\end{equation}
where the time-dependent nonlinearity is given by $U(t)=4\pi a(t)$,
$a(t)$ is the time-dependent two-body scattering
length, and we have taken for simplicity $\hbar=m=1$.

In this work we consider a quasi-one-dimensional Bose-Einstein
condensate (BEC) confined in a strongly anisotropic magnetic trap.
In such a setup the dynamics of the condensate is well-known to
follow a one-dimensional version (with rescaled nonlinearity
coefficient) of the three-dimensional GPE (\ref{GPE_first})
\cite{1DGPE}. Now that the system under scrutiny is
one-dimensional in nature, we can consider the usual magnetic,
parabolic, trapping potential of the form
\begin{equation}
V(x)=\frac{\Omega^{2}}{2}x^{2}.\label{trap_eq}
\end{equation}
Let us now use the following trial wave-function (from now on called the
\emph{q}-Gaussian) to describe the state of the condensate close to its
ground state
\begin{equation}
\psi(x,t)=f\left[q(t)\right]\sqrt{N}\left(1-\frac{(1-q(t))x^{2}}{2w(t)^{2}}\right)^{\frac{1}{1-q(t)}}e^{ix^{2}\beta(t)}.\label{trial_wf_eq}
\end{equation}
The $q$-Gaussian ansatz (\ref{trial_wf_eq}) has three, free,
time-dependent parameters $\{w,\beta,q\}$ corresponding,
respectively,  to the width (which is strictly positive at all
times), chirp, and $q$ parameter, where the $q$ parameter captures
the regime of the BEC cloud (see below).
To determine $f\left[q(t)\right]$ one imposes the normalization
condition
\begin{equation}
\int_{D}\left|\psi(x,t)\right|^{2}dx=N,\label{norm_eq}
\end{equation}
where $N$ represents the total number of atoms in the condensate
and

\begin{eqnarray}
D & = & \left[-\frac{\sqrt{2}w(t)}{\sqrt{1-q(t)}}\ ,
\frac{\sqrt{2}w(t)}{\sqrt{1-q(t)}}\right]\label{eq:integration_domain}
\end{eqnarray}
is the domain supporting the \emph{q}-Gaussian ansatz.
Computing the norm (\ref{norm_eq}) one finds
\begin{equation}
f\left[q(t)\right]=\frac{(1-q(t))^{1/4}}{2^{1/4}w(t)^{1/2}\textrm{B}^{1/2}\left(\frac{1}{2},\frac{-3+q(t)}{-1+q(t)}\right)},\label{fq_eq}
\end{equation}
where $\textrm{B}(\cdot,\cdot)$ is the usual Euler beta
function \cite{functii-speciale}.

It is crucial to note the two important limits of the
\emph{q}-Gaussian ansatz: in the $q(t)\rightarrow1$ limit our
ansatz recovers the usual Gaussian ansatz (valid for
low-density condensates), as
\begin{equation}
\lim_{q(t)\rightarrow1}\left(1-\frac{(1-q(t))x^{2}}{2w(t)^{2}}\right)^{\frac{1}{1-q(t)}}=\exp\left(\frac{-x^{2}}{2w(t)^{2}}\right)\label{Gaussian_rec_eq}
\end{equation}and

\begin{equation}
D=(-\infty, \infty),\label{Gaussian_rec_dom }\end{equation}while
the $q(t)\rightarrow-1$ limit recovers the usual parabolic density
profile of the Thomas-Fermi region (valid for high-density
condensates), as
\begin{equation}
\lim_{q(t)\rightarrow-1}\left(1-\frac{(1-q(t))x^{2}}{2w(t)^{2}}\right)^{\frac{1}{1-q(t)}}=\sqrt{1-\frac{x^{2}}{w(t)^{2}}}\label{TFermi_rec_eq
}\end{equation}and
\begin{equation}
D=[-w(t), w(t)].\label{TFermi_rec_dom}\end{equation}

The novelty of this ansatz is that in addition to having the usual
variational parameters $w(t)$ (width of the condensate) and
$\beta(t)$ (chirp), it \emph{also} accounts for the possible
changes of curvature through the $q(t)$ variable. As the above
limits clearly show $q(t)$ is the crucial ingredient for being
able to recover analytically both the low- and the high-density
regime of a BEC. Naturally, the $q$-Gaussian is not the only
function that is able to interpolate between these two limits of a
BEC \cite{Keceli}. The so-called $S_{n}$ function
\begin{equation}
S_{n}(x)=\exp\left(-\sum_{k=1}^{n}\frac{x^{2k}}{k}\right)\label{Sn_eq}
\end{equation}
becomes a Gaussian for $n=1$ and turns into a parabolic profile for
$n\rightarrow\infty$ (and $\left|x\right|<1$), as
\begin{eqnarray}
\nonumber
S_{\infty}(x)&=&\exp\left(-\sum_{k=1}^{\infty}\frac{x^{2k}}{k}\right)\\[2.0ex]
\nonumber
        &=&\exp\left(\ln\left(1-x^{2}\right)\right)=1-x^{2}.
\label{Sn_limit}
\end{eqnarray}
The advantage of using the $q$-Gaussian lies however in the fact
that this function is amenable to analytical computation (and gives
rise to very simple equations describing the dynamics of the condensate),
while other functions (like the $S_{n}(x)$) are not.

Our variational method follows the traditional recipe: we introduce
the $q$-Gaussian trial wave-function in the BEC Lagrangian defined as
\begin{equation}
\textrm{L}(t)=\int_{D}{\cal L}(x,t)dx,\label{BEC_lag}
\end{equation}
where $D$ is given by (\ref{eq:integration_domain}) and the
Lagrangian density is given by
\begin{equation}
{\cal L}(x,t)=\frac{i}{2}\left(\psi\psi_{t}^{*}-\psi^{*}\psi_{t}\right)+\frac{1}{2}\left|\psi_{x}\right|^{2}+V(x)\left|\psi\right|^{2}+\frac{U(t)}{2}\left|\psi\right|^{4}.\label{density_lag}
\end{equation}
The dynamics of the BEC, restricted to our ansatz, is then
determined through the Euler-Lagrange equations
\begin{equation}
\frac{d}{dt}\frac{\partial\textrm{L}}{\partial\dot{y}}
=\frac{\partial\textrm{L}}{\partial y},\label{Euler_Lagrange_eq}
\end{equation}
where $y=\{w,\beta,q\}$. These equations will be the main result
of our paper and, as we will confirm in the following sections,
they are able to accurately describe the dynamics of the
condensate in both the high- and the low-density limit.

\begin{figure}[t]
\begin{center}
\includegraphics[width=8cm]{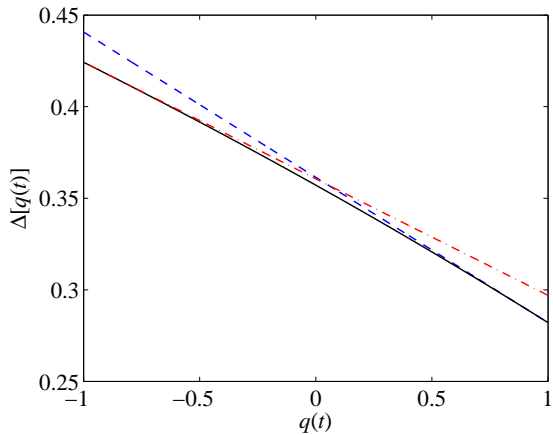}
\end{center}
\caption{(Color online) The solid black line shows
$\Delta\left[q(t)\right]$ as given by Eq.~(\ref{q_func}), while
the blue dashed line and the red dash-dotted line show
$\Delta\left[q(t)\right]$ as given by the approximate formulas
(\ref{Dq_ap_one}) and (\ref{Dq_ap_two}). } \label{fig1}
\end{figure}

A couple of straightforward (though tedious) integrations show
that
\begin{equation}
\frac{\textrm{L}(t)}{N}=\textrm{L}_{1}(t)+\textrm{L}_{2}(t)+\textrm{L}_{3}(t)+\textrm{L}_{4}(t),\label{L_main}
\end{equation}
where
\begin{eqnarray}
\textrm{L}_{1}(t) & = & \frac{2w(t)^{2}\dot{\beta}(t)}{7-3q(t)},\label{L_one}
\\[2.0ex]
\textrm{L}_{2}(t) & = & \frac{NU(t)}{w(t)\sqrt{2}}\Delta\left[q(t)\right],\label{L_two}
\\[2.0ex]
\textrm{L}_{3}(t) & = & \frac{w(t)^{2}\Omega^{2}}{7-3q(t)},\label{L_three}
\\[2.0ex]
\textrm{L}_{4}(t) & = & \frac{5-q(t)}{8w(t)^{2}(1+q(t))}+\frac{4w(t)^{2}\beta(t)^{2}}{7-3q(t)}\label{L_four}
\end{eqnarray}
are the four terms of the BEC Lagrangian. The only apparently cumbersome
term is $\textrm{L}_{2}$ which involves
\begin{eqnarray}
\Delta\left[q(t)\right] & = & \frac{\sqrt{\pi}\Gamma\left(1-\frac{4}{q(t)-1}\right)\sqrt{1-q(t)}}{2\textrm{B}^{2}\left(\frac{1}{2},\frac{q(t)-3}{q(t)-1}\right)\Gamma\left(\frac{3}{2}-\frac{4}{q(t)-1}\right)}.\label{q_func}
\end{eqnarray}
A close numerical inspection of this function shows that it is, however,
\emph{almost} \emph{linear}, as can be easily seen from Fig.~\ref{fig1}.

For practical purposes one can use the two linear approximations:
\begin{equation}
\Delta\left[q(t)\right]  = \frac{41-9q(t)}{64\sqrt{\pi}},\label{Dq_ap_one}
\end{equation}
for the low-density case and
\begin{equation}
\Delta\left[q(t)\right]  =  \frac{51-9q(t)}{100\sqrt{2}},\\[1.0ex]
\label{Dq_ap_two}
\end{equation}
for the high-density one. These linear approximations come from the
series expansion of $\Delta\left[q(t)\right]$ around $q(t)=1$ and
$-1$. It is worth mentioning that in our numerics (see Sec.~\ref{sec:num})
we use the full expression (\ref{q_func}).

Let us notice that by setting $q(t)$ to $-1$ the first term in
$\textrm{L}_{4}$ diverges, a result which is perfectly consistent
with the Thomas-Fermi theory, where one neglects the kinetic term,
which is numerically small in the bulk of the condensate for large
number of atoms, due to its divergence at the border of the cloud.
As will be shown shortly, our variational method works flawlessly
in the high-density limit where $q(t)$ will be close but
\emph{always larger} than $-1$, therefore the divergence will
never be an issue. As far as we know this is the first variational
method that is able to describe dynamically the Thomas-Fermi
limit.

The Euler-Lagrange equations for $\{w, \beta, q\}$ are
\begin{widetext}
\begin{eqnarray}
\frac{q(t)-5}{4w(t)^{3}(1+q(t))}
-\frac{\Delta\left[q(t)\right]NU(t)}{\sqrt{2}w(t)^{2}} +2
w(t)\,\frac{\Omega^{2} +4\beta(t)^{2} +2\dot{\beta}(t)}{7-3q(t)} &
= & 0 \label{main_equations1}
\\
w(t)\left(4\beta(t)-\frac{3\dot{q}(t)}{7-3q(t)}\right) & = &
2\dot{w}(t)
\label{main_equations2}
\\
-\frac{6}{8(1+q(t))^{2}w(t)^{2}}
+\frac{\Delta_{q}\left[q(t)\right]NU(t)}{\sqrt{2}w(t)}
+3w(t)^{2}\, \frac{\Omega^{2} +4\beta(t)^{2}
+2\dot{\beta}(t)}{(7-3q(t))^{2}} & = & 0, \label{main_equations3}
\end{eqnarray}
\end{widetext}
where $(\cdot)_{q}$ stands for derivative with respect to $q$.

Adding an extra exponential term like $\exp(i\phi)$ in the trial
wave-function, \emph{i.e.,} extending the set of variables to
$\{N, \phi, w, \beta, q\}$, gives rise to $\dot{N}=0$, the
equation assuring the conservation of the norm, and
$\dot{\phi}=0$, but leaves the previous equations unchanged. We
have dropped this term for simplicity. Naturally, one can imagine
various other corrections to the present ansatz. However, even
simple phase corrections such as
$\textrm{exp}\left(i\gamma_{n}(t)x^{n}\right)$, where $n$ is an
even integer strictly bigger than two, yield an analytically
intractable Lagrange function.

While Eqs.~(\ref{main_equations1})--(\ref{main_equations3}) are
relatively simple to write down, they are in fact
\emph{differential-algebraic equations} and (if left in their
current form) require specialized numerical treatment (see, for
instance, the monograph of Hairer \emph{et al.} \cite{Hairer}).

To understand the nature of this dynamical system let us remember
that variational methods are usually build around \emph{pairs of
canonically conjugated variables}, which give rise to coupled
ODEs. Taking, for example, the simple case of $q=1$ one easily
notices that $p=w^2/2$ (the generalized momentum) and $q=\beta$
(the generalized coordinate) are the two canonically conjugate
variables. Our ansatz, however, does not posses this intrinsic
symmetry, as $q$ has no canonical conjugate, which finally gives
rise to the algebraic constraint. Since the Lagrange function
becomes analytically intractable at the slightest change in the
ansatz, it is unfruitful to try working with pairs of conjugate
variables. This latter option requires the use of a direct method
\cite{DM}, which has the same lack of transparency as the
numerical solution of the original GPE. Instead, in order to gain
analytical insight, one can easily decouple the previous
equations, a task which is detailed in the next subsection.

\begin{figure}[t]
\begin{center}\includegraphics[width=8cm]{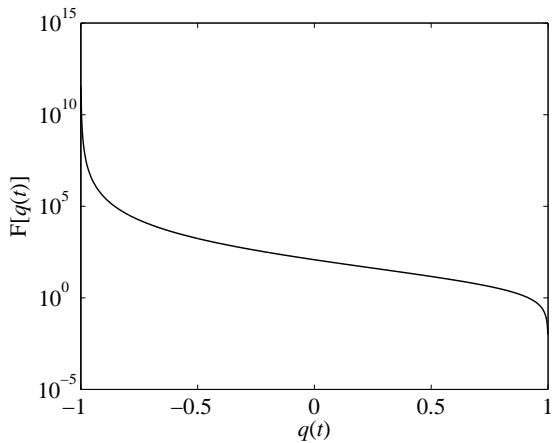}\end{center}
\caption{The line shows $F\left[q(t)\right]$ as given by Eq.~(\ref{F_eq}).
To illustrate the singularity at $q(t)=-1$ we have chosen a log-plot.
Please notice that outside the vicinity of $q(t)=-1$ $F\left[q(t)\right]$
is numerically well behaved. Notice as well that $F\left[q(t)\right]$
has an infinite slope at $q(t)=\pm1$.}
\label{fig2}
\end{figure}

\subsection{Decoupled equations\label{sec:VA_decoup}}

In order to simplify the numerical solution of equations
Eq.~(\ref{main_equations1})-(\ref{main_equations3}) one multiplies
Eq.~(\ref{main_equations1}) by $3w(t)/2(7-3q(t))$ and then
subtracts it from Eq.~(\ref{main_equations2}). The ensuing
equation is purely algebraic. Its solution with respect to $w(t)$
is
\begin{equation}
w(t)=\frac{F\left[q(t)\right]}{NU(t)},\label{main_width}
\end{equation}
where
\begin{eqnarray}
\!\!\!\!\!\!
\!\!\!\!\!\!
{F\left[q(t)\right]} & \equiv &
\displaystyle
\frac{
\displaystyle
\frac{\sqrt{2}}{(1+q(t))}\left(\frac{7-3q(t)}{2(1+q(t))}+\frac{q(t)-5}{4}\right)}{\left\{ \frac{2}{3}\Delta_{q}\left[q(t)\right](7-3q(t))+\Delta\left[q(t)\right]\right\} }.
\label{F_eq}
\end{eqnarray}
Equation (\ref{main_width})
can now be used in the other two equations. We have now decoupled
the equation of the width, $w(t)$, from those of $q(t)$ and $\beta(t)$.
A close numerical inspection of $F\left[q(t)\right]$ shows that it
has a singularity at $q(t)=-1$ (which, as mentioned before, follows
physically from the
divergence of the kinetic energy at the border of the cloud), and
is (numerically) well behaved everywhere else as one can easily infer
from Fig.~\ref{fig2}.

The original variational equations (\ref{main_equations1})--(\ref{main_equations3})
can now be recast as
\begin{equation}
\dot{q}(t)=
\frac{\displaystyle 2\frac{\dot{U}(t)}{U(t)}+4\beta(t)}
{\displaystyle 2\frac{F_{q}\left[q(t)\right]}{F\left[q(t)\right]}-\frac{3}{3q(t)-7}},
\label{q_main}\end{equation}
for $q(t)$ and
\begin{equation}
\dot{\beta}(t)=-2\beta^{2}-\frac{\Omega^{2}}{2}+G\left[q(t)\right]N^{4}U(t)^{4},\label{beta_main}\end{equation}
for $\beta(t)$, where \begin{eqnarray*}
G\left[q(t)\right] & \equiv & (5-q(t)+2\sqrt{2}\Delta\left[q(t)\right]F\left[q(t)\right](1+q(t)))\\
 &  & \times(7-3q(t))F^{-4}\left[q(t)\right](1+q(t))^{-1}/16\\[1.0ex]
 & = & \left\{ 3\Delta\left[q(t)\right]+2\Delta_{q}\left[q(t)\right](7-3q(t))\right\} ^{3}(7-3q)^{2}\\
 &  & \times\left\{ 3\Delta\left[q(t)\right]-\Delta\left[q(t)\right](q(t)-5)(1+q(t))\right\} \\
 &  & \times\, 8(1+q(t))^{6}(q(t)-9)^{-4}(q(t)-1)^{-4}/81.
\end{eqnarray*}
Equations (\ref{q_main}) and (\ref{beta_main}) constitute the main
theoretical result of our work.

\begin{figure}[t]
\begin{center}\includegraphics[width=8cm]{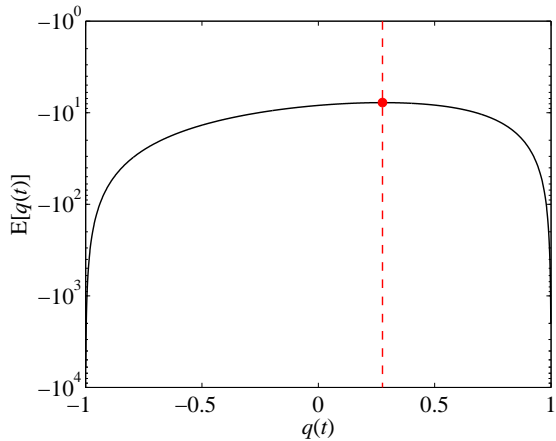}\end{center}
\caption{(Color online) The line shows $E\left[q(t)\right]$ The divergence
around $q(t)=\pm1$ indicates that the low- and the high-density regimes
have a very weak surface dynamics, \emph{i.e.}, $\dot{q}(t)\approx0$.
The red dot (of coordinates $(q_{c},E[q_{c}])=(0.27605...,-7.76722...)$)
and the accompanying dashed red line (of equation $q=q_{c}=0.27605$)
indicate the extremum value of $E\left[q(t)\right]$ at $q_{c}$. }
\end{figure}

Before going into the high- and low-density limits of these two equations
let us make the following comment: the dynamics of $q(t)$ depends
only \emph{indirectly} on the number of particles. It is therefore
extremely instructive to analyze the denominator in Eq.~(\ref{q_main})
\begin{equation}
E\left[q(t)\right]=2\frac{F_{q}\left[q(t)\right]}{F\left[q(t)\right]}-\frac{3}{3q(t)-7},\label{Eoft}\end{equation}
 see Fig. 3, and notice that \emph{it diverges} in the vicinities
of $q(t)=\pm1$ and is well behaved everywhere else. Let us notice
that the extremum value of $E\left[q(t)\right]$ corresponds to
$q_{c}\approx0.27605...$. The divergence of $E\left[q(t)\right]$
originates from the infinite slope of $F\left[q(t)\right]$ at
$q(t)=\pm1$ and indicates that $\dot{q}(t)\approx0$ or, in other
words, that in both the low- and the high-density regime the
dynamics of a BEC affects only \emph{superficially} the curvature
of the wave function. This shows in turn why the use of a Gaussian
ansatz in the Thomas-Fermi regime gives rise to quantitatively
good results (see, for instance, Ref.~\cite{Zoller-var}). The main
point we want to stress is that these two \emph{physically
different} regimes have in common a very weak surface dynamics,
therefore almost any ansatz with an intrinsic breathing mode
provides decent quantitative results.

Finally, let us notice that the Jacobi matrix of the dynamical
system made out of Eqs.~(\ref{q_main}) and (\ref{beta_main})
(computed with the ground state values of $q$ and $\beta$ and, of
course, constant $U$) has complex conjugate roots throughout the
whole density spectrum.

\section{high-density condensates\label{sec:high}}

To see how the previous equations recover the usual hydrodynamic
formulation let us consider for $\Delta[q(t)]$ the approximation
given by Eq. (\ref{Dq_ap_two}). Within this approximation we have
that

\begin{widetext}
\begin{equation}
\dot{q}(t)=\left(2\frac{\dot{U}(t)}{U(t)}+4\beta(t)\right)\left[\frac{3}{7-3q(t)}+\frac{2}{q(t)-9}+\frac{2}{q(t)-1}-\frac{6}{1+q(t)}\right]^{-1}
\label{TFavo}
\end{equation}
and \begin{equation}
\dot{\beta}(t)=-\beta(t)^{2}-\frac{\Omega^{2}}{2}+\frac{2187}{5}\frac{(q(t)-4)(q(t)-3)(1+q(t))^{9}(3q(t)-7)^{2}}{10^{7}\left(9-10q(t)+q(t)^{2}\right)^{4}}N^{4}U(t)^{4}.
\label{TFavt}
\end{equation}

\end{widetext}
As $F[q(t)]$ reduces to
\begin{equation}
F[q(t)]=\frac{50(q(t)-9)(q(t)-1)}{9(1+q(t))^{2}}, \label{F_TF_app}
\end{equation}
and we are investigating the dynamics of the system in a vicinity
of $q(t)=-1$, Eqs. (\ref{TFavo}) and (\ref{TFavt}) take the
simpler form

\begin{equation}
\dot{q}(t)=-\left(2\frac{\dot{U}(t)}{U(t)}+4\beta(t)\right)\left[\frac{9}{10}+\frac{6}{1+q(t)}\right]^{-1}
\label{TFavth}
\end{equation}
 and

\begin{equation}
\dot{\beta}(t)=-\beta(t)^{2}-\frac{\Omega^{2}}{2}+\frac{3NU(t)}{4w(t)^{3}}.
\label{TFavf}
\end{equation}

Equation (\ref{TFavf}) along with the Eq.~(\ref{main_equations2}),
where we now ignore the time derivative of $q(t)$,
\emph{i.e.},\begin{eqnarray} \dot{w}(t) & = &
2w(t)\beta(t),\label{TF_int_four}\end{eqnarray} are the so-called
hydrodynamic equations. To see this more clearly let us work in
the variables $\alpha(t)$ and $a(t)$ defined as
\begin{equation}
\beta(t)=\frac{\alpha(t)}{2}\label{TF_int_five}\end{equation} and

\begin{equation}
a(t)=-\frac{3N}{4w(t)^{3}},\label{TF_int_six}\end{equation} so
that equations (\ref{TFavo}) and (\ref{TFavt}) read
\begin{eqnarray}
\dot{a}(t)+3\alpha(t)a(t) & = & 0\nonumber \\
\dot{\alpha}(t)+\alpha(t)^{2}+\Omega^{2}+2U(t)a(t) & = & 0.\label{TF_final}\end{eqnarray}
These are exactly the hydrodynamic equations put forward by Dalfovo
\emph{et al}. \cite{dalfovo_one}.

\medskip

\section{low-density condensates\label{sec:low}}

\subsection{Gaussian ansatz}

In the low-density limit one can set $q(t)$ to $1$ and ignore its
derivative with respect to time, in which case
Eqs.~(\ref{main_equations1})--(\ref{main_equations3})
can be recast as

\begin{eqnarray}
-\frac{1}{2w(t)^{3}}-\frac{NU(t)}{2\sqrt{2\pi}w(t)^{2}}+\frac{\Omega^{2}w(t)}{2}\nonumber \\
+2w(t)\beta(t)^{2}+w(t)\dot{\beta}(t) & = & 0 \nonumber\\
2w(t)\beta(t) & = & \dot{w}(t).\label{main_equations_GL}\end{eqnarray}

From the second equation in (\ref{main_equations_GL}) one has that

\begin{eqnarray}
\ddot{w}(t) & = & 4w(t)\beta(t)^{2}+2w(t)\dot{\beta}(t),\label{GL_int_one}\end{eqnarray}
therefore \begin{equation}
\ddot{w}(t)-\frac{1}{w(t)^{3}}-\frac{NU(t)}{\sqrt{2\pi}w(t)^{2}}+\Omega^{2}w(t)=0.\label{GL_final}\end{equation}
This is the well-known equation that describe the dynamics of the
width of the condensate in the low-density limit.

\subsection{\emph{q}-Gaussian ansatz}

One way to refine the standard equation stemming from the Gaussian
ansatz is to start from Eqs.~(\ref{q_main}) and (\ref{beta_main})
and use for $\Delta\left[q(t)\right]$ the high-density
approximation provided by Eq. (\ref{Dq_ap_one}). The ensuing
equations are

\begin{widetext}

\begin{equation}
\dot{q}(t)=\left(2\frac{\dot{U}(t)}{U(t)}+4\beta(t)\right)\left[\frac{2}{q(t)-9}+\frac{2}{q(t)-1}-\frac{4}{1+q(t)}+\frac{3}{7-3q(t)}+\frac{18}{1-9q(t)}\right]^{-1}
\label{qGvo}
\end{equation}

and

\begin{equation}
\dot{\beta}(t)=-\beta^{2}-\frac{\Omega^{2}}{2}+\frac{(7-3q(t))^{2}(1+q(t))^{6}(9q(t)-1)^{3}(26+3(q(t)-7)q(t))}{2097152\pi^{2}(q(t)-9)^{4}(q(t)-1)^{4}}N^{4}U(t)^{4}.
\label{qGvd}
\end{equation}
\end{widetext}

As $F[q(t)]$ reduces to
\begin{equation}
F[q(t)]=\frac{16\sqrt{2\pi}(q(t)-9)(q(t)-1)}{(1+q(t))^2(9q(t)-1)},
\label{FqGapp}
\end{equation}
and we are investigating the dynamics of the system in a vicinity
of $q(t)=1$, Eqs. (\ref{qGvo}) and (\ref{qGvd}) take the simpler
form

\begin{equation}
\dot{q}(t)=\frac{q(t)-1}{2}\left(2\frac{\dot{U}(t)}{U(t)}+4\beta(t)\right)
 \label{qGvth}
\end{equation}and
\begin{equation}
\dot{\beta}(t)=-\beta(t)^{2}-\frac{\Omega^{2}}{2}-\frac{1}{2w(t)^4}-\frac{N
U(t)}{u(q)\sqrt{2\pi}w(t)^3}, \label{qGvf}
\end{equation}where
\begin{equation}
u(q)=\frac{128}{142-81q(t)+27q(t)^2.} \label{u_form}
\end{equation}

Replacing in Eq. (\ref{FqGapp}) $q(t)$ with 1 in all factors
except the one yielding a zero term, one has that

\begin{equation}
q(t)=1-\frac{N U(t)w(t)}{4\sqrt{2\pi}}. \label{qconec1}
\end{equation} 
Finally, inserting $q(t)$ in Eq. (\ref{qGvth}) one has
that
\begin{equation}
\dot{w}(t)=2w(t)\beta(t). \label{qconec2}
\end{equation}

Except for the $u(q)$ factor Eqs. (\ref{qGvf}) and (\ref{qconec2})
are identical to Eqs. (\ref{main_equations_GL}). Of course, the
two sets of equations coincide in the $N=0$ limit, but otherwise
the ones stemming from the \emph{q}-Gaussian ansatz have the clear
advantage of accounting for the change in the curvature of the
wave-function.

\begin{figure}[htbp]
\begin{center}
\includegraphics[width=4.3cm,height=4.5cm]{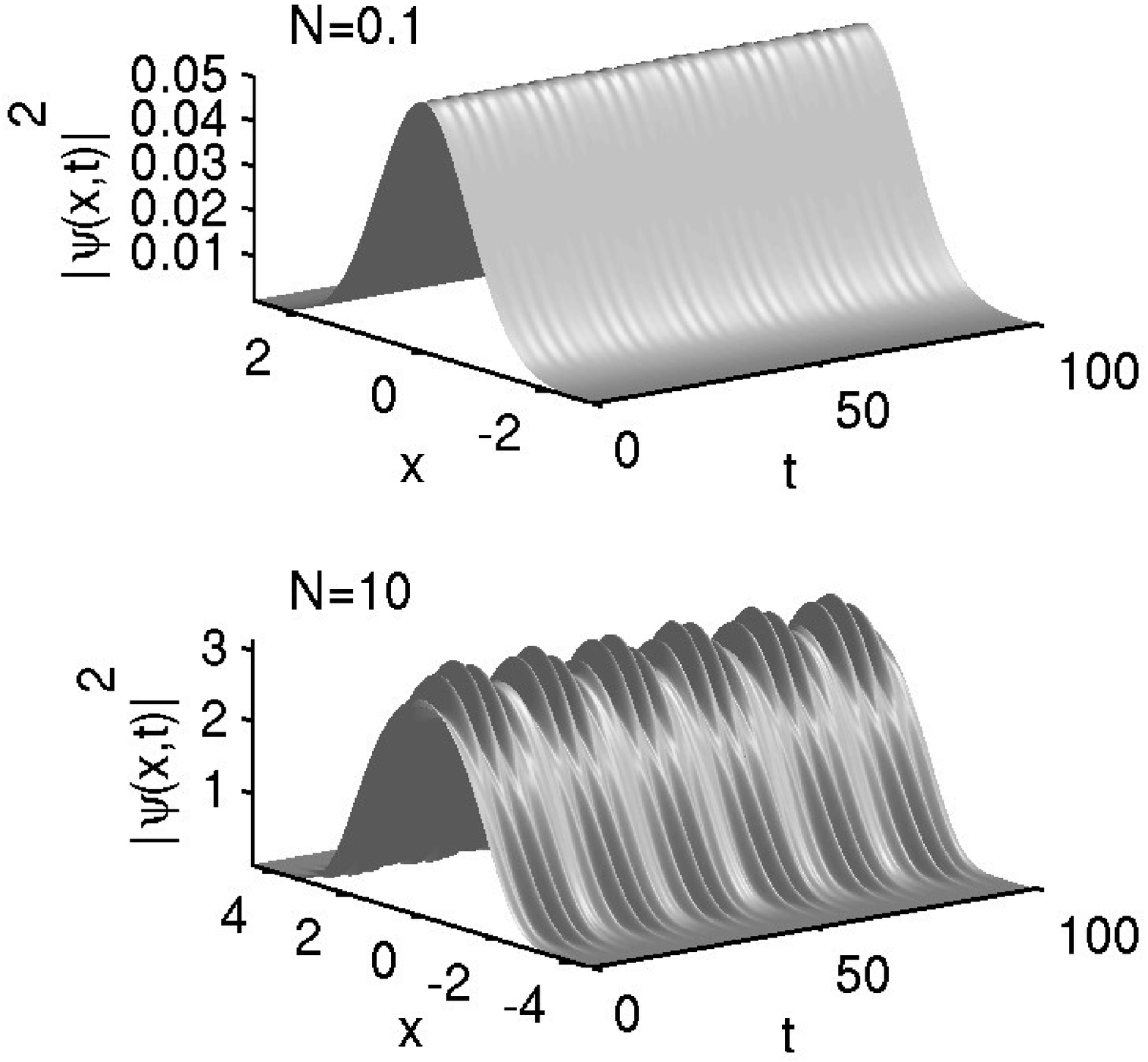}
\hskip-0.1cm
\includegraphics[width=4.3cm,height=4.5cm]{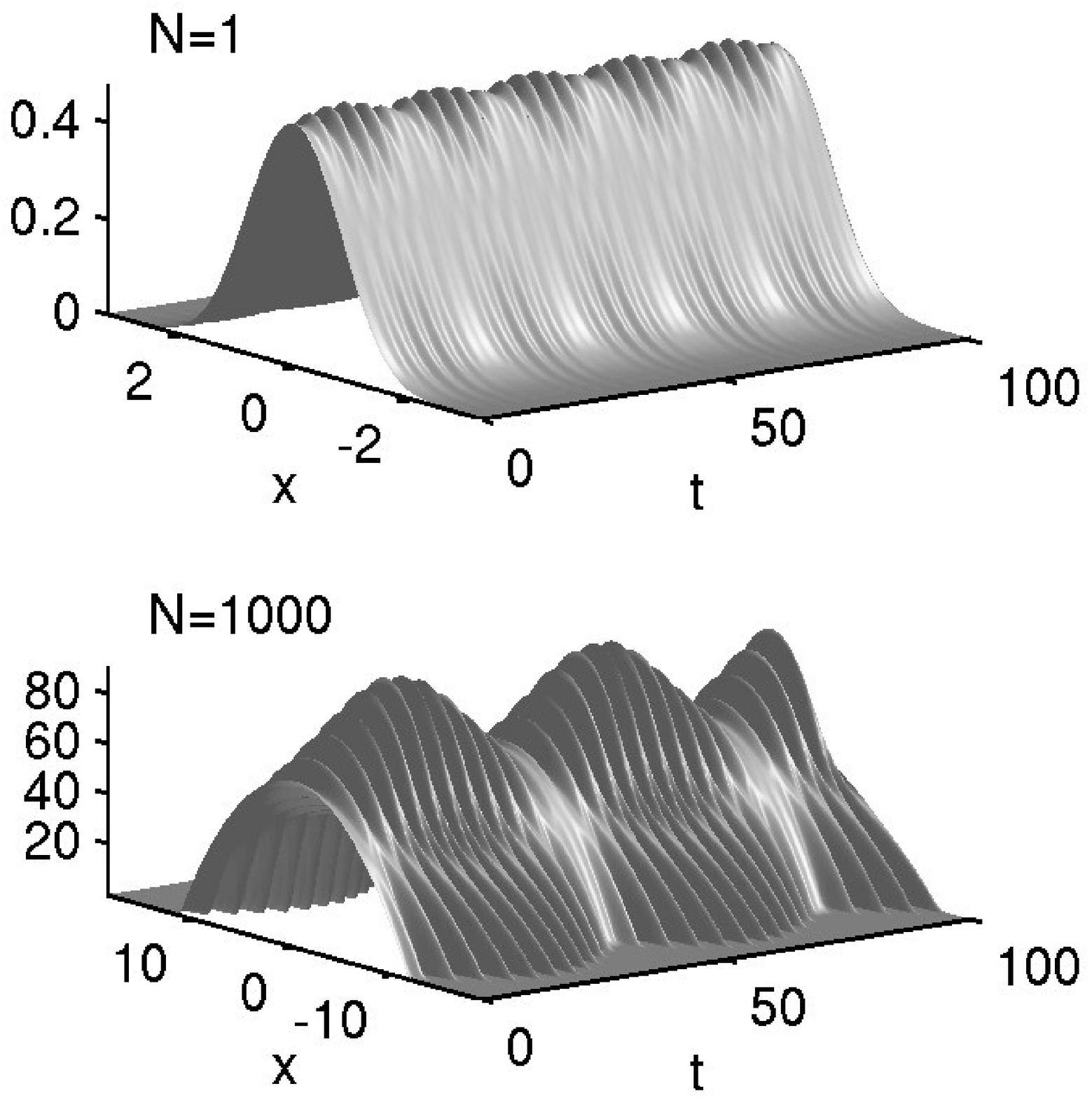}
\end{center}

\caption{Dynamics of the BEC density profile under the modulation
of the nonlinearity for different atom numbers $N$ (as indicated
on each plot). Parameter values correspond to: $\Omega=1$,
$U_{0}=1$, $\epsilon=0.1$ and $\omega=1.6$.} \label{plot_allu}
\end{figure}

\section{numerical results\label{sec:num}}

In the previous two sections we have shown that the equations
stemming from the \emph{q}-Gaussian ansatz recover those of the
usual Gaussian approach (in the low-density regime) and the
standard high-density hydrodynamic equations. We now turn to a
detailed numerical comparison between our variational equations
and the original GPE. To this end, we consider a magnetic trap of
the form (\ref{trap_eq}) and a modulated nonlinear term
$U(t)=U_{0}(1+\epsilon\sin(\omega t))$. The modulation of the
nonlinear term can be achieved by modulation of the transverse
width (see Ref.~\cite{faraday} and references therein) or by, the
so-called, Feshbach resonance technique
\cite{feshbach-mag,feshbach-opt}.

\begin{figure}[htbp]
\begin{center}
\hskip0.15cm\includegraphics[width=8.3cm]{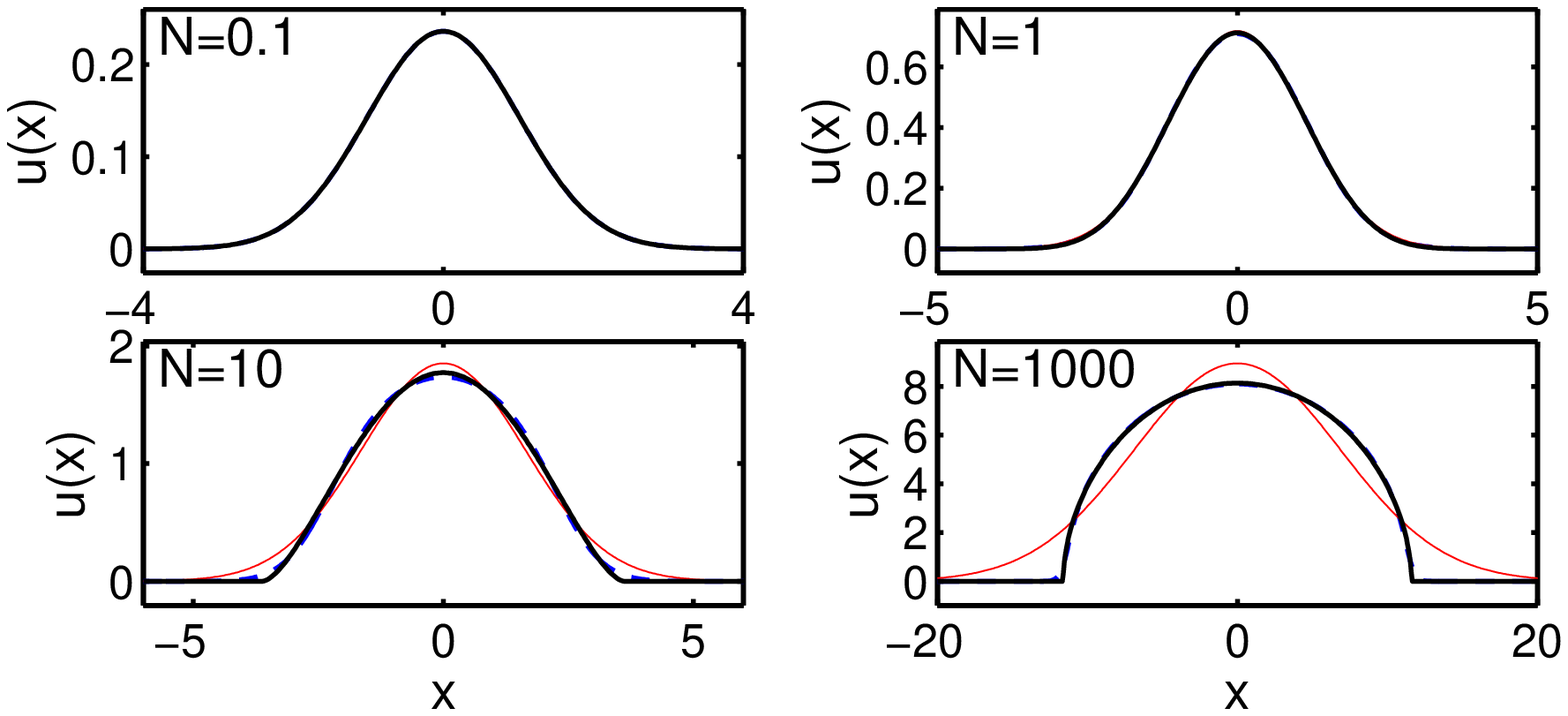}
\end{center}
\vspace{-0.7cm}
\begin{center}
\includegraphics[width=8.5cm]{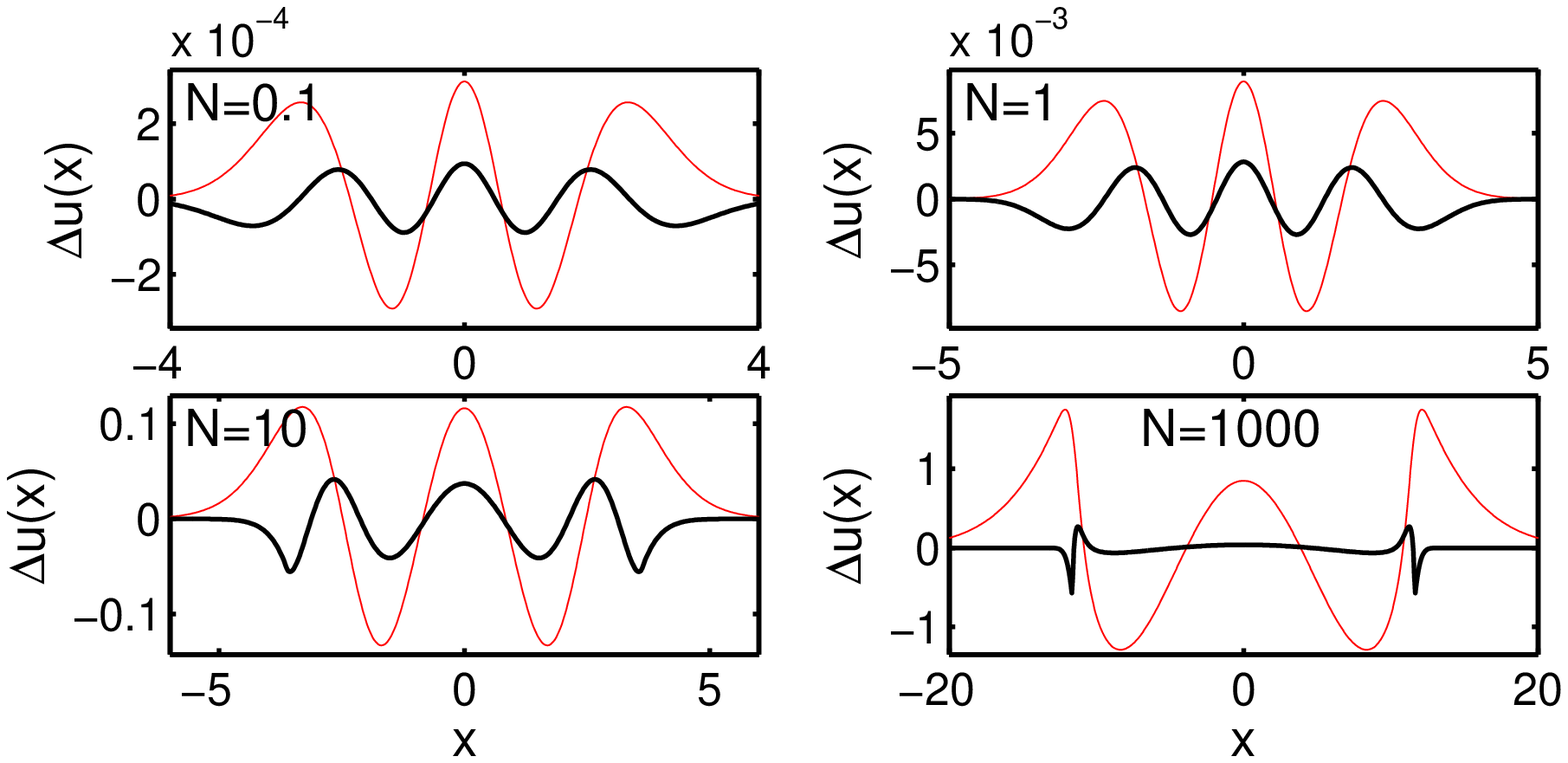}
\end{center}
\vspace{-0.5cm} \caption{(Color online) Comparison between the GPE
steady state solution, the \emph{q}-Gaussian fit and the Gaussian
fit. The top four panels show the steady state solution to the GPE
(thick black dashed line), its fit using the \emph{q}-Gaussian
(thick black solid line, almost indistinguishable from the dashed
line due to the very good approximation) and the traditional
Gaussian (thin red solid line) for different atom numbers $N$. The
bottom four panels show the respective differences between the
steady state and the \emph{q}-Gaussian (thick black line) and the
steady state and the traditional Gaussian (thin red line). }
\label{plot_uini_g}
\end{figure}

For illustration purposes we choose a magnetic trap with $\Omega=1$
and a nonlinearity modulation with
$U_{0}=1$, $\epsilon=0.1$ and $\omega=1.6$.
Let us notice that the natural frequency of the condensate is
equal to $\Omega\omega_{N}$, where $\omega_{N}$ depends weakly on
$N$ and is numerically close to $2$, therefore choosing
$\omega=1.6$ assures a series of beats (or, technically,
parametric resonances) between the frequency of the driving field
and the natural frequency of the condensate as it can be observed
in the space-time plots of the density for different values of $N$
in Fig.~\ref{plot_allu} (for a more detailed analysis of
mode-locking and energy transfer due to resonances see
Refs.~\cite{ours1} and \cite{ours2} and references therein).

\begin{figure}[htbp]
\begin{center}
\hskip0.20cm\includegraphics[width=8.3cm]{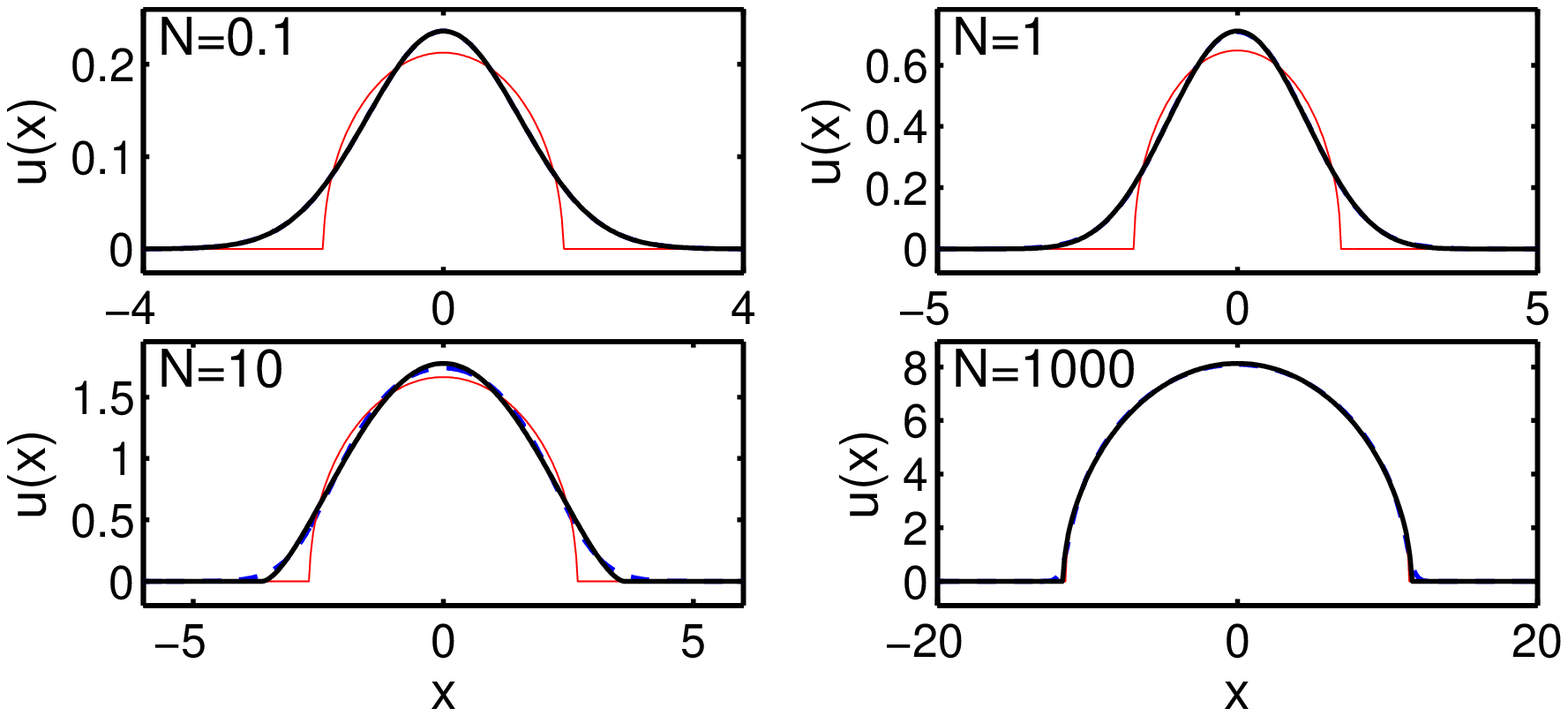}
\end{center}
\vspace{-0.6cm}
\begin{center}
\includegraphics[width=8.5cm]{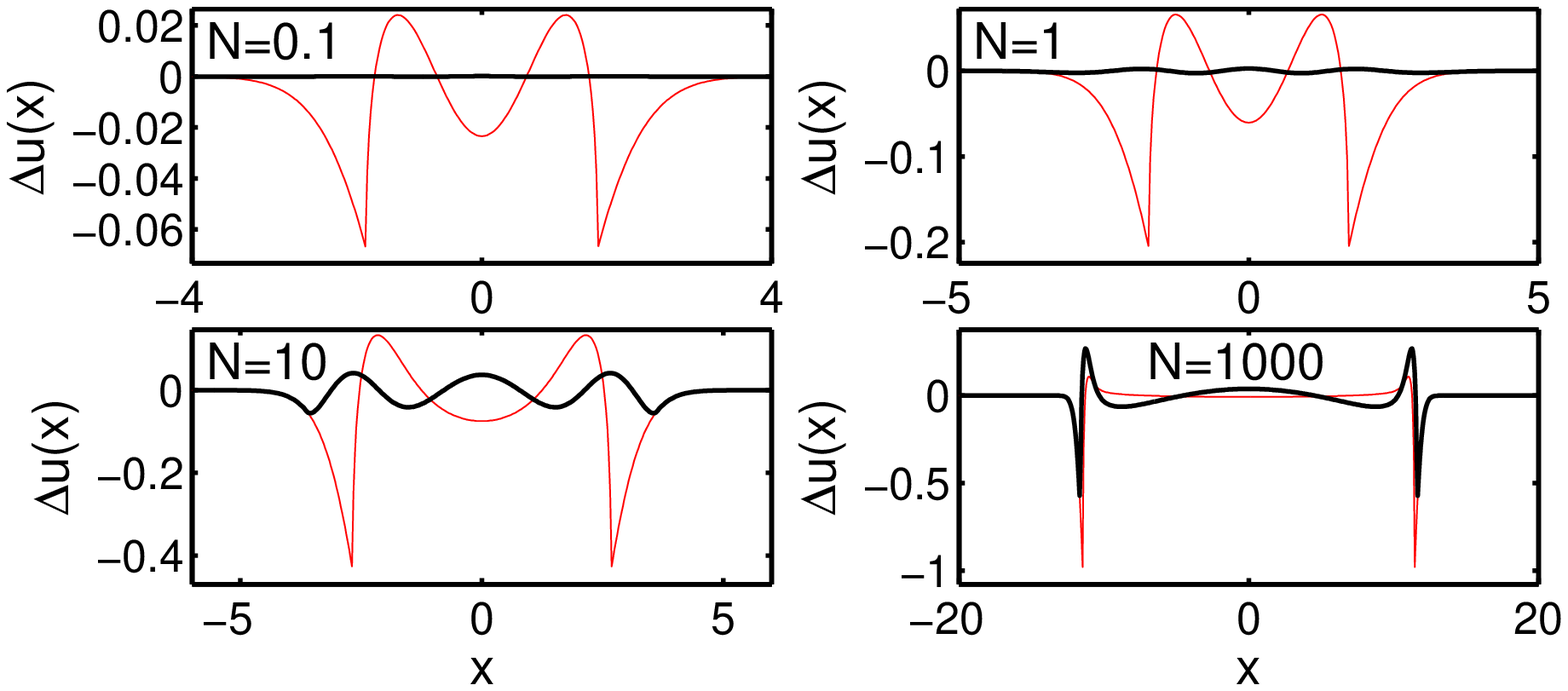}
\end{center}
\vspace{-0.5cm}
\caption{(Color online) Comparison between the GPE steady state
solution, the \emph{q}-Gaussian fit and the parabolic Thomas-Fermi
profile fit. The panels show the same information as in
Fig.~\ref{plot_uini_g}. by replacing the traditional Gaussian by
the parabolic Thomas-Fermi profile. } \label{plot_uini_p}
\end{figure}

The advantage of our approach is that close to resonance a system
responds strongest to excitations, therefore this is the ideal
setup to investigate the limits of our equations.
In order to probe the different regimes of the condensate we
choose five different cases corresponding to
$N=0.1,\:1,\:10,\:10^{2},\:10^{3}$. These cases encompass the bulk
part of the density spectrum, the corresponding values of \emph{q}
(restricted to only three significant digits) being equal to
$0.990$, $0.906$, $0.431$, $-0.236$, and $-0.710$, respectively.
In Figs.~\ref{plot_uini_g} and \ref{plot_uini_p} we depict, for
various values of $N$, the steady state solution $u(x)$ to the GPE
alongside the best linear square fits using the \emph{q}-Gaussian,
the traditional Gaussian (see Fig.~\ref{plot_uini_g}) and the
parabolic Thomas-Fermi approximation (see Fig.~\ref{plot_uini_p}).
The steady state $u(x)$ of Eq.~(\ref{GPE_first}) is obtained by
taking
$$\psi(x,t)=u(x) e^{-i \mu t}$$
where $\mu$ is the
so-called chemical potential. The ensuing steady state equation
is then solved using a fixed point iteration algorithm (Newton
method) by optimizing $\mu$ to give the desired total mass.
In Figs.~\ref{plot_uini_g} and \ref{plot_uini_p}
we also depict (see lower set of panels) $\Delta
u(x)=u_{\rm GPE}(x)-u_{\rm fit}(x)$, \emph{i.e.}, the difference
between the actual steady state of the GPE and the corresponding
fits. As it can be clearly seen from both figures, the
\emph{q}-Gaussian provides an extremely good approximation of the
wave-function in \emph{both} the low- and the high-density
regimes. In fact, as it can be evidence from
Fig.~\ref{plot_uini_g}, even in the low-energy regime, where it is
well known that the traditional Gaussian ansatz is a good
approximation, the \emph{q}-Gaussian is superior. Incidentally the
same is true when analyzing the high-density regime where, as it
can be evidence from Fig.~\ref{plot_uini_p}, the \emph{q}-Gaussian
provides a better approximation than the Thomas-Fermi profile.

\begin{figure}[htbp]
\begin{center}
\includegraphics[width=8.6cm,height=4.9cm]{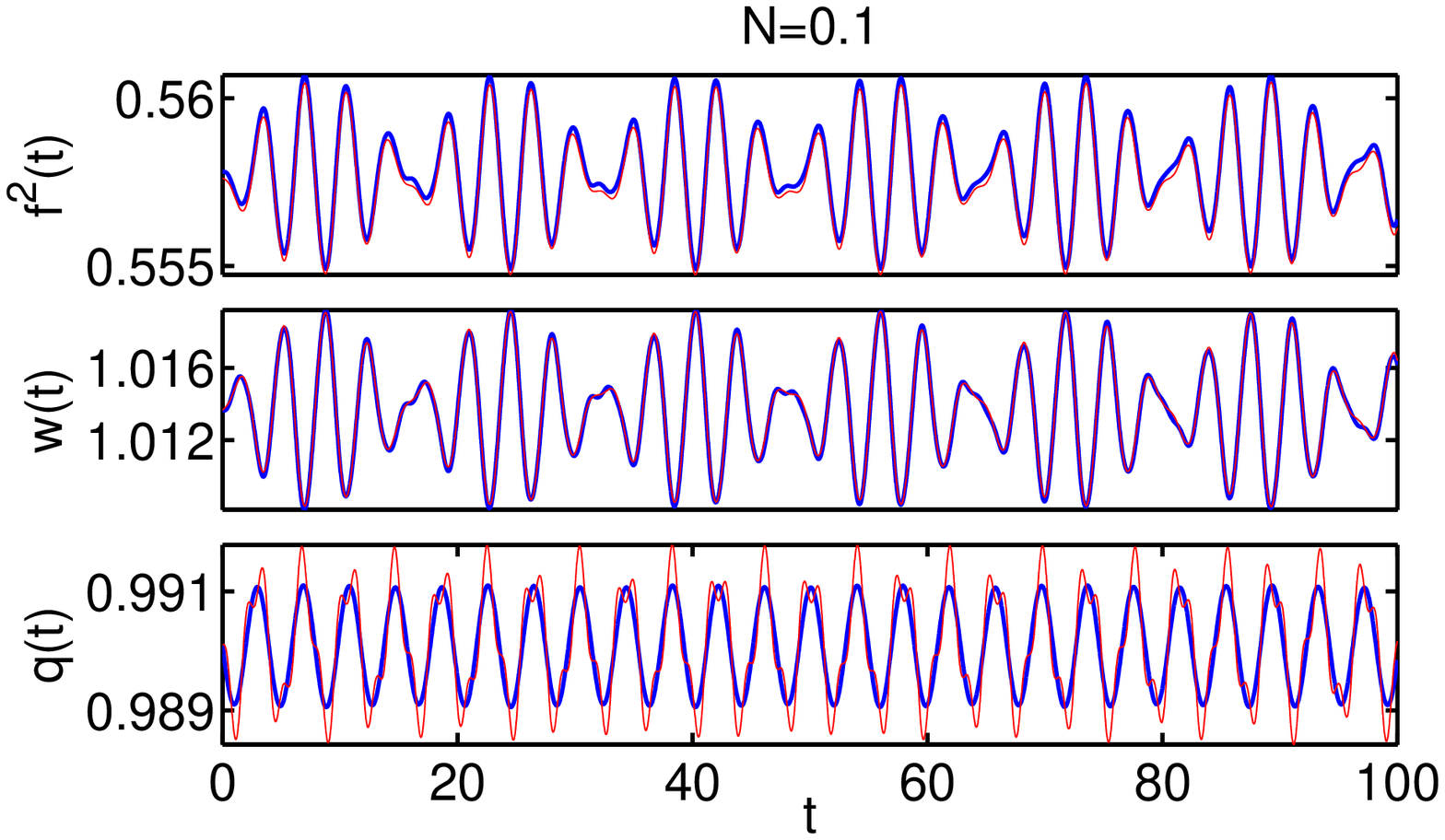}~~
\end{center}
\vspace{-0.4cm}
\begin{center}
\includegraphics[width=8.6cm,height=4.9cm]{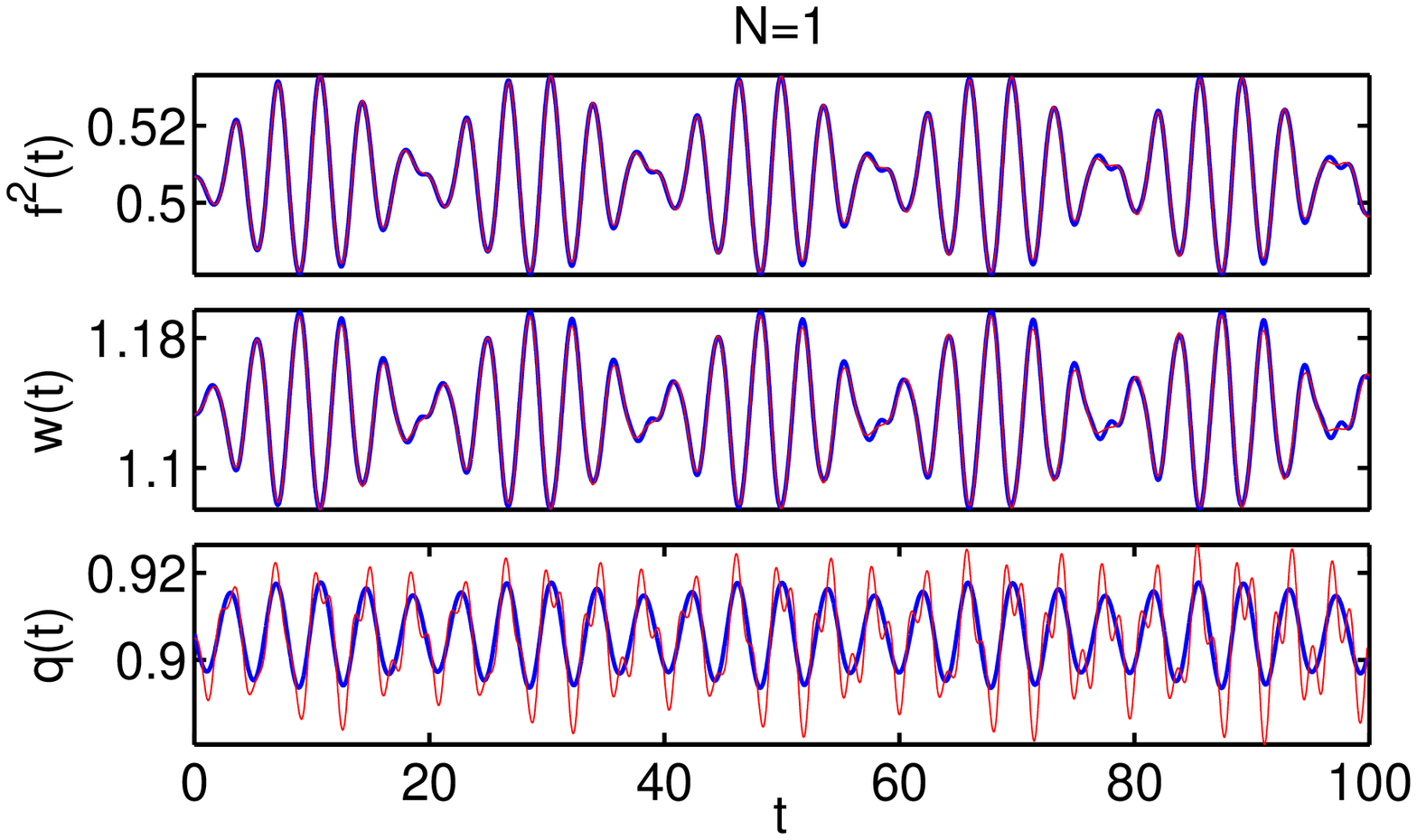}
\end{center}
\vspace{-0.4cm}
\caption{(Color online) Low-density regime. Comparison between the
full GPE dynamics and the reduced ODE dynamics using the
\emph{q}-Gaussian approach. The top (bottom) series of panels
corresponds to atom numbers $N=0.1$ ($N=1$). For each atom number,
the three panels corresponds to, from top to bottom, the time-series
of a) the peak density $f^2[q(t)]$, b) width $w(t)$, and c)
the \emph{q}-Gaussian parameter $q(t)$. Thick blue lines
corresponds to the full GPE dynamics while thin red lines to the
reduced ODE dynamics using the \emph{q}-Gaussian variational
approach. Note that for $f^2[q(t)]$ and $w(t)$ the time-series
obtained from the full GPE dynamics and our ODE reduction
are practically indistinguishable.
} \label{qGauss1}
\end{figure}

\begin{figure}[ht]
\begin{center}
\includegraphics[width=8.6cm,height=4.9cm]{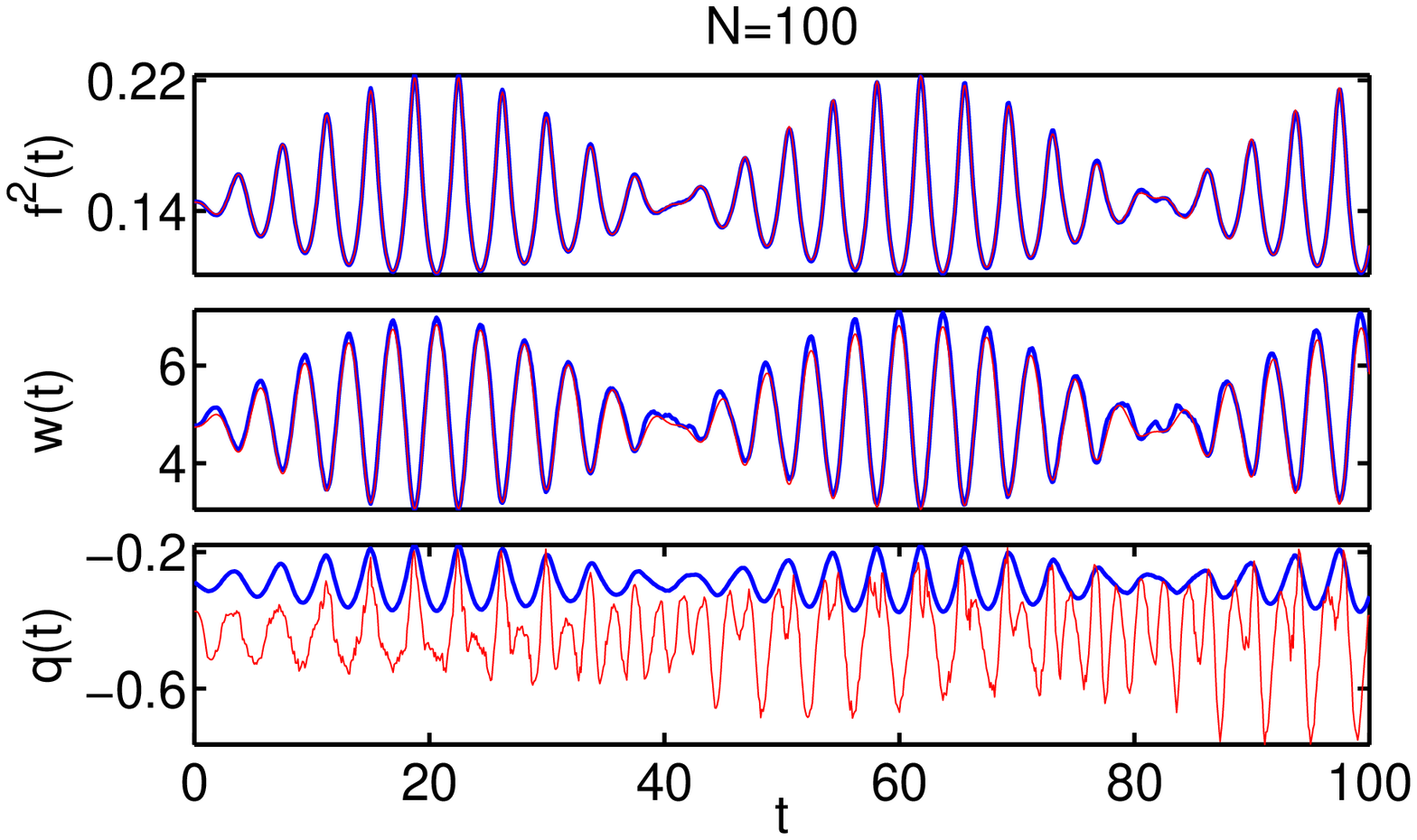}
\end{center}
\vspace{-0.6cm}
\begin{center}
\includegraphics[width=8.6cm,height=4.9cm]{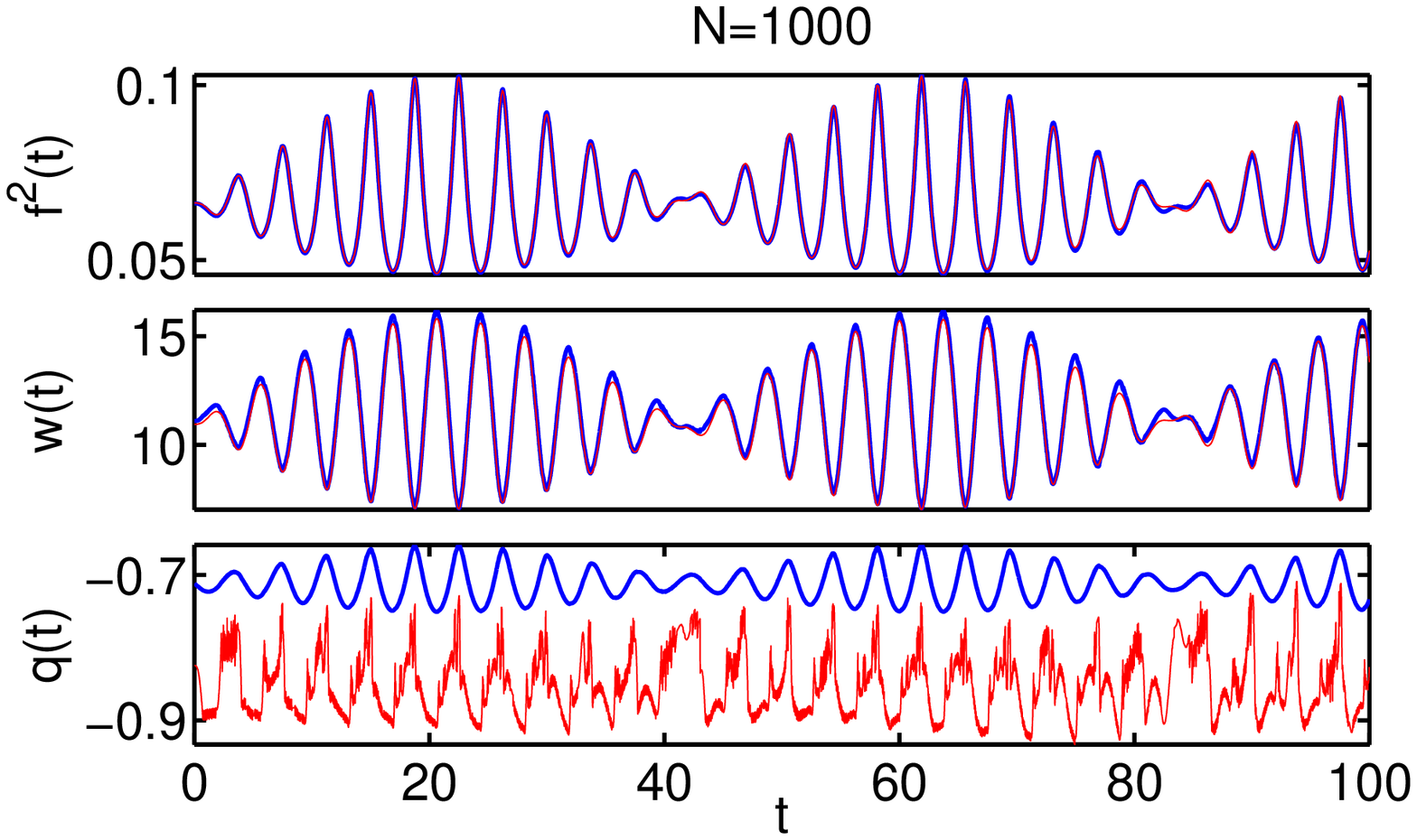}
\end{center}
\vspace{-0.4cm}
\caption{(Color online) High-density regime. All
panels depict same information as in Fig.~\ref{qGauss1} but in
this case for $N=10^2$ and $N=10^3$. } \label{qGauss2}
\end{figure}

\begin{figure}[ht]
\begin{center}
\includegraphics[width=8.6cm,height=4.9cm]{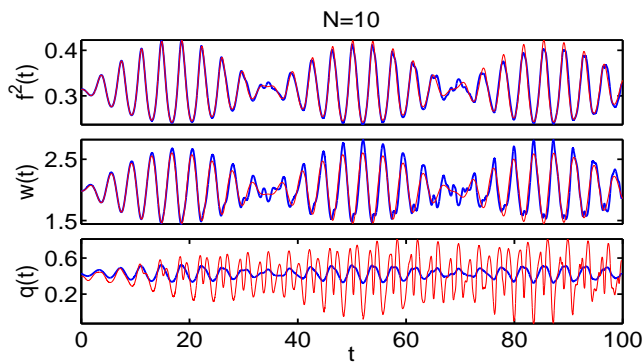}
\end{center}
\vspace{-0.4cm}
\caption{(Color online) Intermediate density regime. All panels
depict same information as in Fig.~\ref{qGauss1} but in this case
for $N=10$. } \label{qGauss3}
\end{figure}

\begin{figure}[ht]
\begin{center}
\includegraphics[width=8.8cm,height=7.4cm]{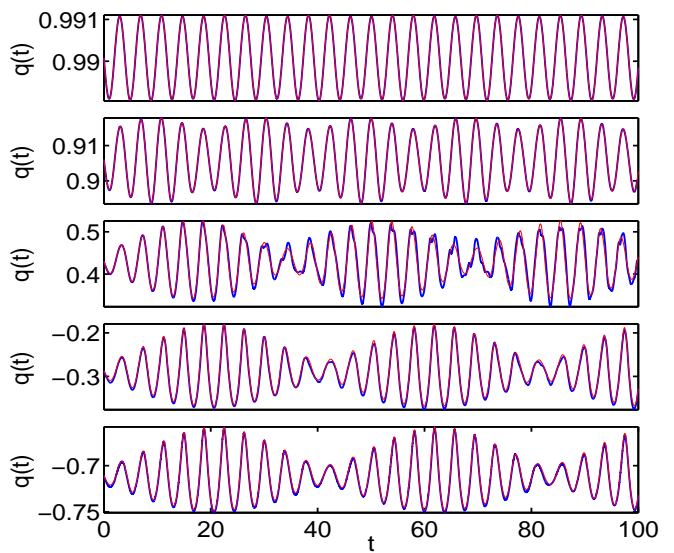}
\end{center}
\vspace{-0.5cm}
\caption{(Color online)
Comparison of the \emph{q}-Gaussian parameter $q(t)$
between full GPE dynamics (thick blue line) and the reduced
ODE dynamics by fitting $q(t)$ from $w(t)$ (thin red line).
Each panel, from top to bottom, corresponds to:
a) $N=10^{-1}$,
b) $N=10^{0}$,
c) $N=10^{2}$,
d) $N=10^{3}$, and
e) $N=10^{4}$.
}
\label{qfits}
\end{figure}

\begin{figure}[htbp]
\begin{center}
\includegraphics[width=8.6cm,height=7.4cm]{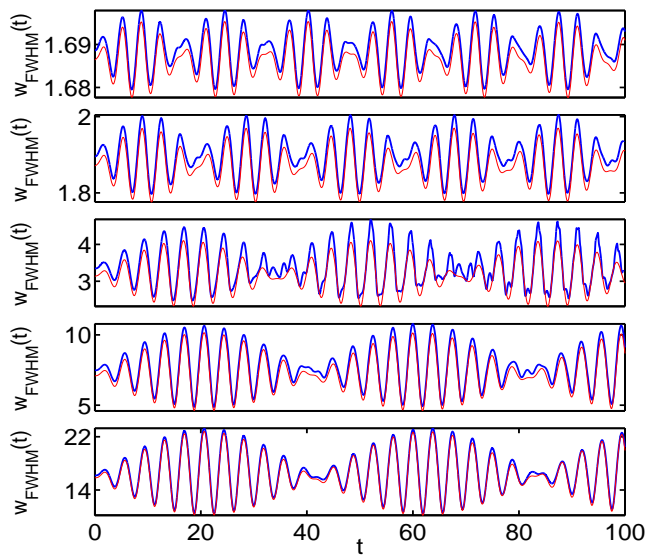}
\end{center}
\vspace{-0.5cm}
\caption{(Color online)
Full width at half maximum $w_{\rm FWHM}(t)$.
Each panel, from top to bottom, corresponds to:
a) $N=10^{-1}$,
b) $N=10^{0}$,
c) $N=10^{2}$,
d) $N=10^{3}$, and
e) $N=10^{4}$.
Thick blue lines corresponds to the full GPE dynamics
while thin red lines to the reduced ODE dynamics
using the \emph{q}-Gaussian variational approach.
}
\label{fwhm}
\end{figure}

\begin{figure}[ht]
\begin{center}
\includegraphics[width=8.8cm,height=7.4cm]{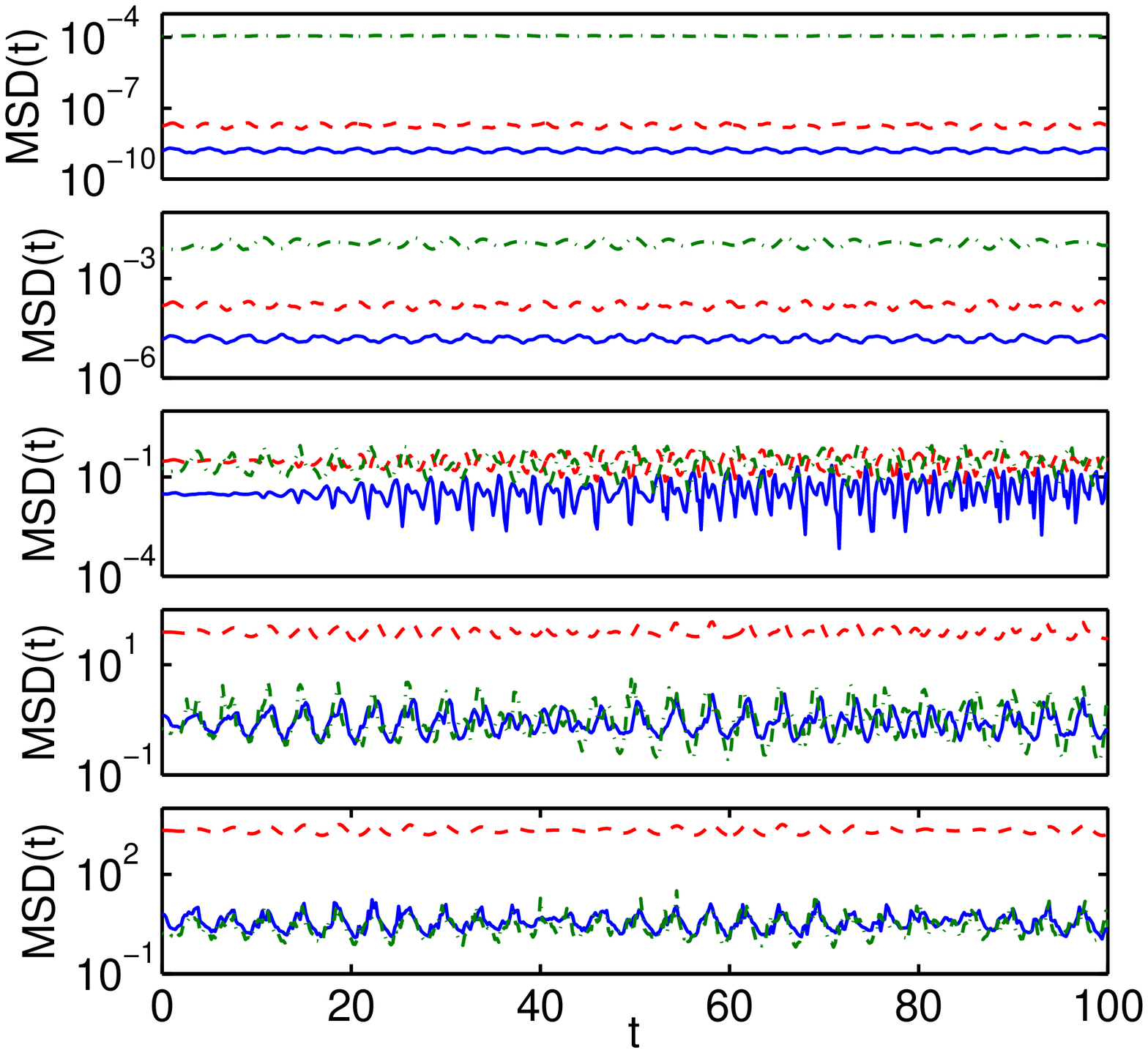}
\end{center}
\vspace{-0.5cm}
\caption{(Color online)
Mean square deviation (MSD) between the densities profiles
obtained from the different fits and the density
obtained from the original GPE.
The blue solid line corresponds to the \emph{q}-Gaussian approach,
while the red dashed line corresponds to the traditional Gaussian
and the green dash-dotted corresponds to the TF approximation.
As in Fig.~\ref{qfits},
each panel, from top to bottom, corresponds to:
a) $N=10^{-1}$,
b) $N=10^{0}$,
c) $N=10^{2}$,
d) $N=10^{3}$, and
e) $N=10^{4}$.
}
\label{msd}
\end{figure}

To quantify the accuracy of our variational treatment of the GPE we
will follow two distinct veins: \emph{i.}) we fit the spatial component
of the time-dependent density profile provided by a GPE solver with
a \emph{q}-Gaussian profile and compare the ensuing time-series for
$f^2\left[q(t)\right]$, $w(t)$, and $q(t)$ with the corresponding
ones obtained from Eqs.~(\ref{main_equations1})--(\ref{main_equations3});
also, we compute
the full width at half maximum (FWHM), $w_{\rm FWHM}(t)$,
as a function of time both for
the density profile provided by the GPE solver and that obtained from
Eqs.~(\ref{main_equations1})--(\ref{main_equations3});
\emph{ii}.) using two different fits
based on a Gaussian and a Thomas-Fermi density profile, and the \emph{q}-Gaussian
profile that yields from Eqs.~(\ref{main_equations1})--(\ref{main_equations3}),
we calculate
the time-dependent mean square deviation (MSD) of these profiles from
that obtained from the original GPE.

Figs.~\ref{qGauss1}, \ref{qGauss2}, and \ref{qGauss3} depict the
$f^2\left[q(t)\right]$, $w(t)$, and $q(t)$ time-series for,
respectively, the low-density regime ($N=0.1,1$), the high-density
regime ($N=10^2,10^3$), and the intermediate density regime
($N=10$). Notice the very good agreement for
$f^2\left[q(t)\right]$ and $w(t)$ for all the cases under
consideration.

In the low density limit, when the condensate cloud is close to a
Gaussian profile, one can observe from Fig.~\ref{qGauss1} that the
dynamics of the parameter $q$ for the full GPE numerics and the
reduced ODE system are in good agreement. However, in the high-
and intermediate density limits, as it can be evidenced from
Figs.~\ref{qGauss2} and \ref{qGauss3}, there seems to be a
(somewhat counterintuitive) poor agreement in the $q$ time-series
despite the fact that $f^2\left[q(t)\right]$ and $w(t)$ are still
in very good agreement. This apparent shortcoming should not be seen as
a fault of our variational treatment, but rather as an indication
of the fitting sensitivity of the \emph{q}-component of our
density profile. In fact, we have observed that even the addition
of small quantities of noise ($<5\%$) to the profile to be fitted
induce sizeable changes in the resulting $q$ value (results
not shown here). To understand
this more clearly one should look into how the fitting
procedure works: when minimizing the mean-square deviation between
the GPE profile and the \emph{q}-Gaussian one encounters the
derivative of the \emph{q}-Gaussian with respect to \emph{q} which
has a log-type of singularity at the border of the cloud. This
singularity makes it very hard to make a one-to-one comparison of
$q(t)$. One way to avoid this problem is to compute $q(t)$ (via
Eq.~(\ref{main_width})) from the $w(t)$. This latter approach (see
Fig.~\ref{qfits}) shows good agreement between the two
time-series.

Another way to test the quality of the \emph{q}-Gaussian ansatz is
to make a direct comparison between the  $w_{\rm FWHM}(t)$
time-series (see Fig.~\ref{fwhm}). The $w_{\rm FWHM}(t)$ of the
GPE density profile is computed numerically, while that of the
\emph{q}-Gaussian is given by

\begin{equation}
w_{\rm
FWHM}(t)=\frac{2\sqrt{-2+2^{1/2(1+q(t))}}}{\sqrt{-1+q(t)}}w(t).\label{wFWHMqGaussian}\end{equation}
Please notice the trivial limits $q(t)\rightarrow 1$ and
$q(t)\rightarrow -1$, in which cases we get the well-known results
$w_{\rm FWHM}(t)=2\sqrt{\log{2}}w(t)$ and $w_{\rm
FWHM}(t)=\sqrt{2}w(t)$.

In order to further substantiate the quality of the
\emph{q}-Gaussian approach we depict in Fig.~\ref{msd} the mean
square deviation between the different fits (note the logarithmic
scale in the vertical axis). In the low-density regime the
\emph{q}-Gaussian function performs some four orders of magnitude
better than the Thomas-Fermi fit. This result is rather obvious
since the Thomas-Fermi limit is not the correct limit to use in
this case. However, when comparing the \emph{q}-Gaussian (blue
solid line) with the traditional Gaussian (red dashed line) one
can still notice more than one order of magnitude better agreement
for the \emph{q}-Gaussian trial function. On the other extreme of
the regime (\emph{i.e.}, the high-density regime), the
\emph{q}-Gaussian performs about two orders of magnitude better
than the traditional Gaussian. This is again obvious since the
traditional Gaussian is not the right limit to use in this regime.
On the other hand, it is interesting to note that the
\emph{q}-Gaussian has a similar performance to the Thomas-Fermi
approximation in this high-density regime. Finally, in the
interesting, and more challenging, case of intermediate density
$N=10$, where the condensed cloud profile is neither close to
Gaussian nor parabolic, the \emph{q}-Gaussian clearly outperforms
\emph{both} the traditional Gaussian and the parabolic
Thomas-Fermi approximation by about one order of magnitude.

Therefore, the results presented above, are clear evidence that
not only the \emph{q}-Gaussian is able to provide a good
variational ODE reduction of the full GPE across \emph{all}
the different density regimes, but it is also has a better
(or equal) performance than the separate, traditional,
approximations at either extreme of the density regime.

\section{conclusions\label{sec:conclu}}

In this paper we have proposed an efficient variational method to
investigate the spatio-temporal dynamics of magnetically-trapped
Bose-condensed gases. To this end we have employed a
\emph{q}-Gaussian trial wave-function that interpolates between
the low- and the high-density limit of the ground state of a
Bose-condensed gas. Our main result consists of reducing the
Gross-Pitaevskii equation to a set of only three equations:
\emph{two coupled nonlinear ordinary differential equations}
describing the phase and the curvature of the wave-function and
\emph{a separate algebraic equation} yielding the generalized
width. On the analytical side our equations recover those of the
usual Gaussian variational approach (in the low-density limit),
and the hydrodynamic equations that describe the high-density
case. On the numerical side, we have presented a detailed
comparison between the numerical results of our reduced equations
and those of the original GPE for the case of periodically driven
BEC.
We used as indicators of the \emph{q}-Gaussian performance the
fitting of the peak density, the condensate width, and $q(t)$, a
direct comparison between the full width at half maximum obtained
from our variational treatment and that of the original equation,
and the mean square deviation calculations. All these indicators
demonstrate that the \emph{q}-Gaussian variational reduction
performs very well over all the density regimes. Furthermore, we
found that a) in the low-density regime, the \emph{q}-Gaussian
clearly outperforms the traditional Gaussian; b) in the
high-density limit the \emph{q}-Gaussian has a similar performance
to the Thomas-Fermi approximation; c) and that in the intermediate
density regime the \emph{q}-Gaussian clearly outperforms
\emph{both} the traditional Gaussian and the Thomas-Fermi
approximation.

Finally, let us mention that while our equations are
one-dimensional they can be easily extended to two- and
three-dimensional clouds. This avenue is currently under
investigation and will be reported elsewhere.

\begin{acknowledgments}
The authors thank Mogens H.~Jensen, Henrik Smith, Christopher
Pethick and Poul Olesen for fruitful discussions on the subject.
A.I.N.~acknowledges the hospitality of San Diego State University
where part of this work was carried out. R.C.G.~acknowledges
support from NSF-DMS-0505663.
\end{acknowledgments}

\end{document}